# Advancing Audio Fingerprinting Accuracy with AI and ML: Addressing Background Noise and Distortion Challenges

1st Navin Kamuni
*AI ML M.Tech*
*BITS Pilani WILP*
USA
navin.kamuni@gmail.com

2nd Sathishkumar Chintala
*Department of IT*
*Fisher Investments*
USA
sathishkumarchintala9@gmail.com

*3rd Naveen Kunchakuri*
*Department of IT*
*Eudoxia Resrach University*
USA
Knav18@gmail.com

*4th Jyothi* Swaroop Arlagadda Narasimharaju
*Google*
*Santa Clara, CA*
anjyothiswaroop@gmail.com

5th Venkat Kumar
AI ML M.Tech
*BITS Pilani WILP*
India
venkatkumarr.vk99@gmail.com

**ABSTRACT** — Audio fingerprinting, exemplified by pioneers like Shazam, has transformed digital audio recognition. However, existing systems struggle with accuracy in challenging conditions, limiting broad applicability. This research proposes an AI and ML integrated audio fingerprinting algorithm to enhance accuracy. Built on the Dejavu Project's foundations, the study emphasizes real-world scenario simulations with diverse background noises and distortions. Signal processing, central to Dejavu's model, includes the Fast Fourier Transform, spectrograms, and peak extraction. The "constellation" concept and fingerprint hashing enable unique song identification. Performance evaluation attests to 100% accuracy within a 5-second audio input, with a system showcasing predictable matching speed for efficiency. Storage analysis highlights the critical space-speed trade-off for practical implementation. This research advances audio fingerprinting's adaptability, addressing challenges in varied environments and applications.

## INTRODUCTION

The emergence of audio fingerprinting technology has revolutionized digital audio recognition[1]. Applications like Shazam have played a pivotal role in seamlessly identifying songs from extensive libraries, providing users with an effortless music discovery experience[2][3]. However, despite the marvel of these technologies, the high accuracy demonstrated in optimal conditions often falters in the face of challenges such as background noise, poor audio quality, or signal distortions[4]–[9][10]. This limitation inhibits the widespread applicability of current audio fingerprinting systems in diverse real-world scenarios, including live music events, outdoor environments, and user-generated content on digital platforms[7], [11]–[13]. To address these challenges, this research aims to pioneer an advanced audio fingerprinting algorithm that integrates Artificial Intelligence (AI) and Machine Learning (ML) techniques. This algorithm seeks to effectively filter and compensate for background noise and distortions, enhancing the robustness and reliability of audio fingerprinting across a broader range of environments and applications.

Inspired by the user experience of applications like Shazam, this research aims to create a technological marvel capable of recognizing songs from extensive audio databases. Beyond Shazam, the exploration of audio fingerprinting includes projects like SoundHound, Midomi, Chromaprint, and Echoprint, showcasing diverse approaches to solving the intricate puzzle of song identification from audio data. The foundation of this research is built upon the Dejavu Project[14], an open-source audio fingerprinting initiative implemented in Python and licensed under the MIT License (https://github.com/worldveil/dejavu). This paper explores the theoretical foundations, methodologies, and evaluation metrics used in Dejavu, providing valuable insights into audio signal processing and the principles of fingerprinting.

## DATA COLLECTION AND ANALYSIS: A COMPREHENSIVE ENDEAVOR IN AUDIO SAMPLING

Our research embarked on a pivotal phase in audio fingerprinting—Data Collection and Analysis. This involved meticulously assembling a comprehensive dataset, curated to encompass diverse audio samples with background noises and distortion levels. Recognizing that real-world scenarios seldom offer pristine, noise-free audio environments, our dataset became a microcosm of auditory landscapes, featuring an amalgamation of environmental sounds, distortions, and nuances. The data collection process followed a methodical approach, ensuring the dataset's richness and diversity. A wide array of audio samples spanning genres, languages, and recording conditions was meticulously sourced. These samples captured the auditory diversity of daily life—from the subtle hums of urban environments to potential distortions in crowded spaces. Background noises were deliberately introduced to simulate scenarios requiring the audio fingerprinting system to discern signals amidst noise. Ambient sounds, chatter, traffic noises, and other environmental factors were layered into the audio samples, creating a complex auditory backdrop. Distortion levels added another dimension to our dataset, simulating less-than-ideal conditions. Audio samples underwent controlled distortion, ranging from subtle variations to deliberate alterations in audio quality. Our analysis aimed to gain insights into how background noises and distortion levels impacted the audio fingerprinting system, covering quantitative metrics and qualitative assessments[6], [10], [14]–[16].

## MODEL DEVELOPMENT

Signal processing is fundamental to Dejavu's model, playing a pivotal role in recognizing audio signals amidst noise and distortions. The subsequent sections delve into key signal processing concepts and techniques employed in Dejavu's architecture.

### 1. Audio as a Signal

At the core of Dejavu's model is the representation of audio as a signal. In signal processing, music is digitally encoded as a sequence of numbers. An uncompressed .wav file's richness, with 44,100 samples per channel per second, underscores the information contained in audio signals.The Fast Fourier Transform (FFT), a crucial tool, efficiently analyzes music data.

Samples_per_sec = Duration × Samples_per_sec_per_channel (1)

Here, Samples_per_sec represents the total number of samples per second, Duration is the audio length, and Samples_per_sec_per_channel denotes samples per second per channel.

### 2. Sampling and Nyquist-Shannon Sampling Theorem

The choice of 44,100 samples per second adheres to the Nyquist-Shannon Sampling Theorem, ensuring accurate representation of frequencies below half the sampling rate. Given that humans can't hear frequencies above 20,000 Hz, a sampling rate of 44,100 Hz is deemed sufficient[17][18]

$$Samples_per_sec_needed = Highest_Frequency \times 2 \qquad (2)$$

### 3. Spectrograms and FFT

Spectrograms, visualizing signal amplitudes at different frequencies over time, are crucial in signal processing. Created through repeated FFT application, spectrograms provide a 2D array depicting amplitude as a function of time and frequency. Challenges arise in real-world scenarios where background noise introduces complexities. Dejavu utilizes FFT over small time windows to create a spectrogram, representing amplitude over time and frequency (see Figure 1) [14].

$$Spectrogram(t,f) = FFT(Samples(t,f)) \qquad (3)$$

Here, Spectrogram(t,f) denotes the spectrogram at time t and frequency f, and Samples(t,f) represents audio samples at time t and frequency f.

Spectrograms are generated using the Short-Time Fourier Transform (STFT), calculated as:

$$STFT(t,f) = \sum_{n=0}^{N-1} Samples(t+n) \cdot e^{-j2\pi fn} \qquad (4)$$

Where $STFT(t,f)$ is the STFT at time t and frequency f, $Samples(t+n)$ are the audio samples in the time window, and $e^{-j2\pi fn}$ represents the complex exponential.

### 4. Peak Extraction

Peak extraction involves identifying instances where the amplitude significantly surpasses surrounding values, defined by:

$$\begin{cases} STFT(t,f) \\ 0 \end{cases}$$

*if the STFT value is greater than a predefined than a predefined Thereshold otherwise* (5)

Here, the Threshold is a predefined value distinguishing peaks from background noise. For each moment in time (t) and frequency (f), we assess if the Short-Time Fourier Transform (STFT(t,f)) surpasses a predefined threshold. If affirmative, we designate it as a peak (Peak(t,f)); otherwise, it is considered background noise, set to zero. This equation aids in discerning crucial peaks in the audio signal, effectively isolating them from surrounding noise through a predefined threshold (see Figure 2) [14].

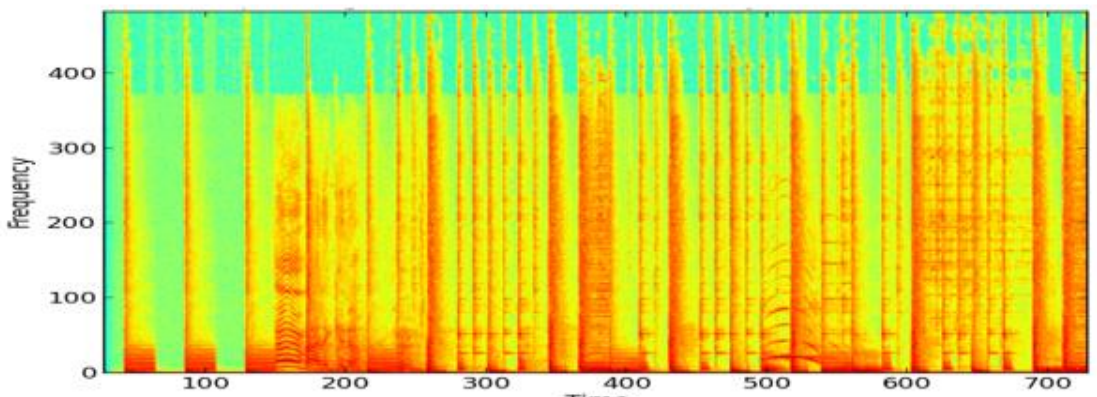

**Figure 1:** presents a spectrogram of the first few seconds of "Blurred Lines" by Robin Thicke. Spectrograms condense the frequency and amplitude information of an audio signal into a visual form, making it easier for analysis[14].

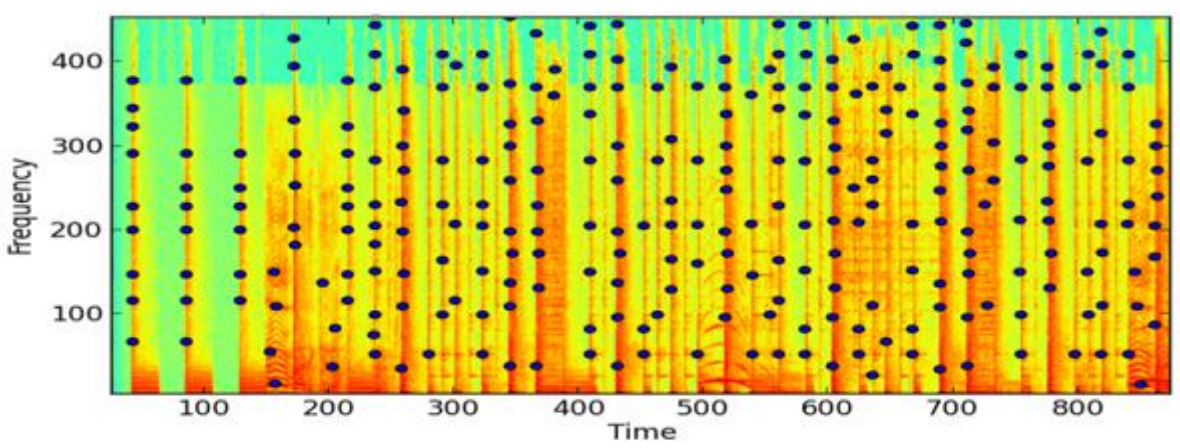

**Figure 2**: presents the spectrogram of "Blurred Lines" after the extraction of noise-resistant peaks. These peaks, serving as pivotal points of interest, essentially formulate a distinctive "fingerprint" for the song[14].

## 5. Constellation of Peaks

In alignment with the Shazam whitepaper's analogy, the grouping of peaks into a "constellation" serves as a pivotal method for song identification. **Figure 3** meticulously presents a magnified and annotated spectrogram snippet, providing a visual emphasis on this intricate concept[14]. The intricate process of forming a constellation of peaks entails the strategic pairing of peaks, considering the time delta between them:

$$Constellation(P1, P2, \Delta t) = (f1, f2, \Delta t) \quad (6)$$

Here, P1(t1,f1) and P2(t2,f2) denote two peaks, and $\Delta t = t2 - t1$ signifies the time delta. This triplet encapsulates not only the frequency information of the paired peaks but also their temporal separation in a comprehensive manner.

## 6. Fingerprint Hashing

Following the constellation extraction process, the subsequent stage involves transforming these constellations into fingerprints through a meticulous hashing technique. Fingerprint hashing, in this context, employs a hash function on the frequencies of peaks and their temporal differences

$$Hash(P1, P2, \Delta t) = Hash(f1, f2, \Delta t) \quad (7)$$

This hashing mechanism functions as a succinct representation of the celestial constellation, facilitating efficient storage and retrieval processes (see Figure 4) [14].

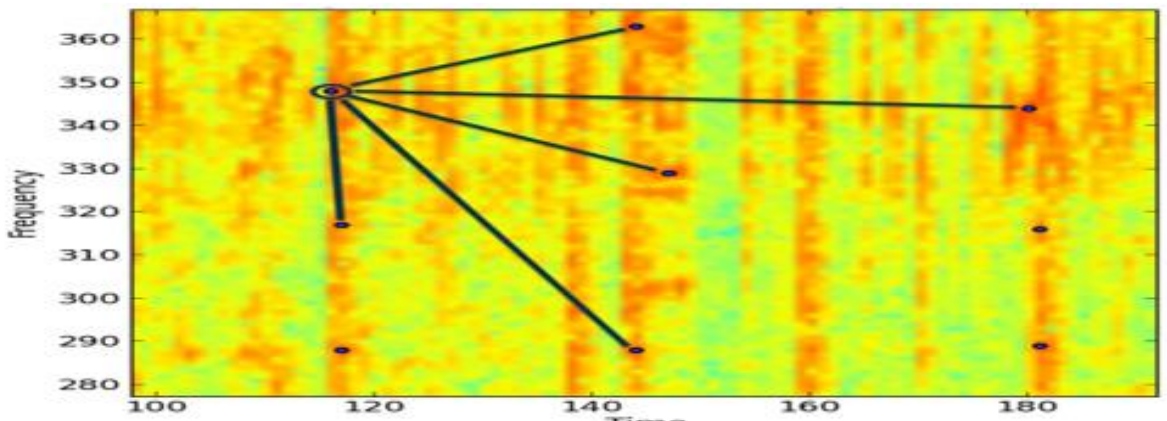

**Figure 3**. Zoomed-in annotated spectrogram snippet presents a magnified and annotated spectrogram snippet, providing a visual emphasis on this intricate concept[14].

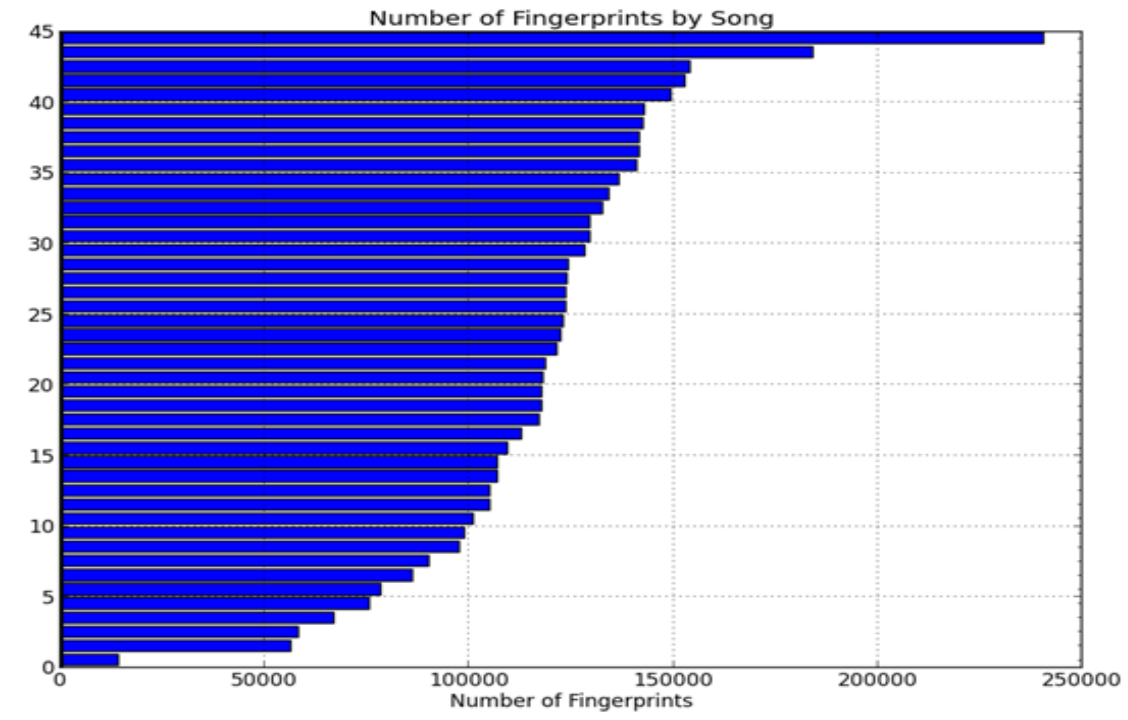

**Figure 4**: provides an informative visual representation, showcasing the quantity of fingerprints associated with each song. This insight into the diversity and abundance of unique fingerprints contributes to a comprehensive understanding [14].

## 7. Song Learning and Recognition

The Dejavu system undertakes two pivotal tasks: learning new songs through fingerprinting and recognizing unknown songs by searching within the learned song database. This process is heavily reliant on the stored fingerprints in the database. Learning a new song involves the insertion of fingerprints into the database, associating them with the corresponding song ID

$$INSERT\ INTO\ fingerprints\ (hash,\ song_{id},\ offset) \quad (8)$$

Here, hash is the fingerprint hash, song_id is the unique identifier for the song, and offset represents the time window from the spectrogram where the hash originated.

Recognizing an unknown song encompasses capturing audio samples, processing them to extract hashes, and matching these hashes with the database:

$$find_database_matches(hashes) \quad (9)$$

The identified matches are aligned based on their time differences to predict the song

## 8. Fingerprint Alignment

Accurate song prediction hinges on the precise alignment of fingerprints. Adjusting the offset, representing the time window from the original track, is vital to match the relative offset from the sample, thereby accommodating variations in the starting point of audio recording.

The adjusted difference is calculated as:

$$adjusted\ difference = database\ offset\ from\ original\ track - sample\ offset\ from\ recording \quad (10)$$

This adjustment ensures that the relative offsets of true matches are consistent, assuming the playback speed aligns with the recording speed.

## 9. Performance Evaluation

Illustrated in Figure 5 is the Dejavu Recognition Accuracy over time, showcasing the system's remarkable precision even with minimal audio input[14]. The corresponding accuracy percentages, detailed in **Table 1** for varying audio durations, underscore Dejavu's commendable accuracy. Notably, with just a one-second audio input, Dejavu achieves 60% accuracy, progressing to 100% with inputs of 5 seconds or more[14].

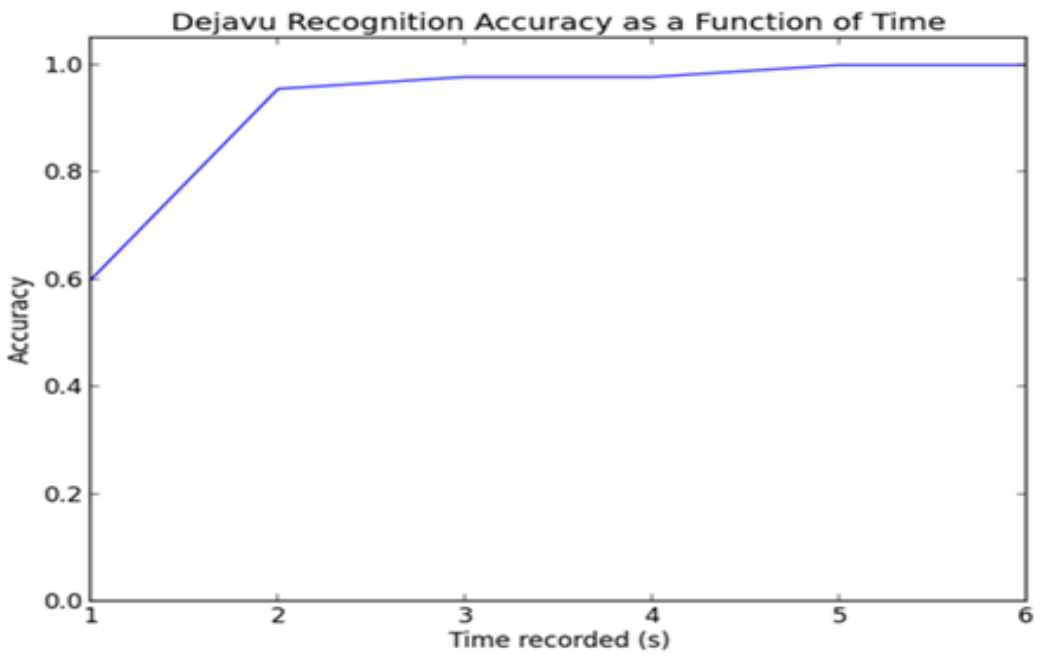

**Figure 5:** Dejavu Recognition Accuracy as a Function of Time[14].

## 10. Performance Metrics: Speed and Storage

The linear relationship depicted in Figure 6 and described by Equation 11 emphasizes the consistent matching speed. Additionally, **Table 2** sheds light on the storage demands for various audio information types, highlighting the substantial space occupied by fingerprints and underscoring the trade-off between storage space and processing speed.

The linear relationship is represented by the equation

$$1.364757 \times record\ time - 0.034373 = time\ to\ match \quad (11)$$

This equation signifies the predictable nature of matching time, largely dependent on the length of the spectrogram created.

**Table 1.** Recognition Accuracy for Different Durations [14].

| Number of Seconds | Number Correct | Percentage Accuracy |
|---|---|---|
| **1** | 27 / 45 | 60.00% |
| **2** | 43 / 45 | 95.60% |
| **3** | 44 / 45 | 97.80% |
| **4** | 44 / 45 | 97.80% |
| **5** | 45 / 45 | 100.00% |
| **6** | 45 / 45 | 100.00% |

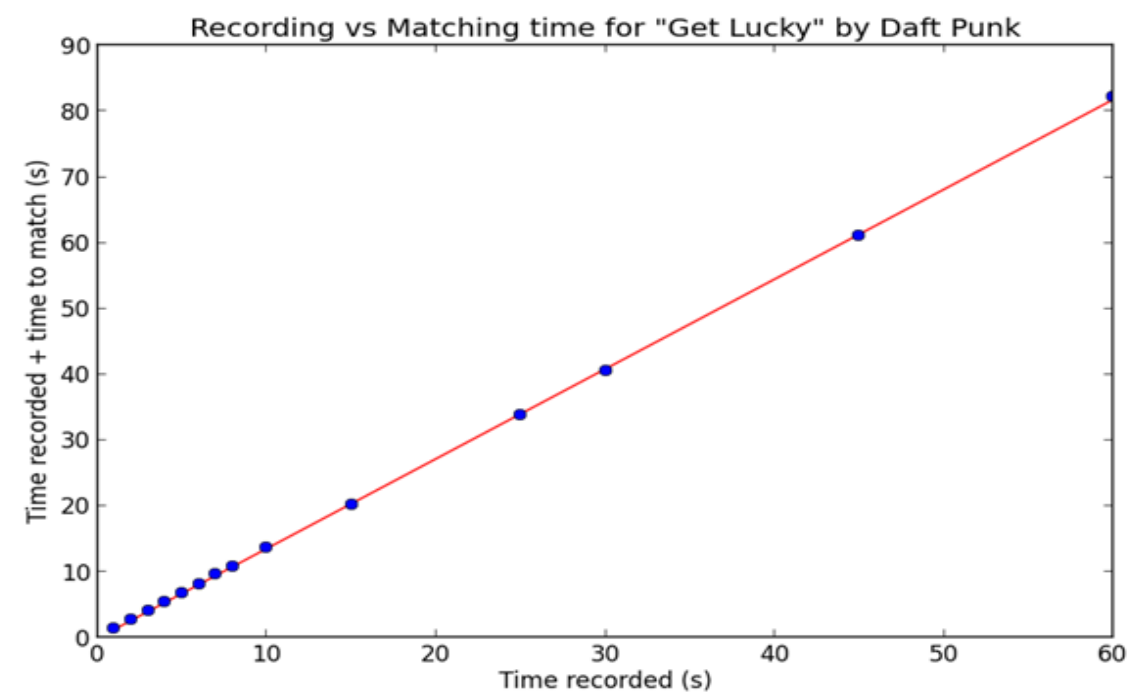

**Figure 6:** illustrates the relationship between recording time and matching time for the song "Get Lucky" by Daft Punk. The linear trend indicates a consistent matching speed, emphasizing the system's efficiency[14].

**Table 2** provides insights into the storage requirements for different audio information types. Notably, fingerprints occupy a substantial amount of space, emphasizing the trade-off between space and speed[14].

| Audio Information Type | Storage in MB |
|---|---|
| **mp3** | 339 |
| **wav** | 1885 |
| **fingerprints** | 377 |

## Conclusion

In conclusion, this research represents a significant leap forward in audio fingerprinting, leveraging the capabilities of Artificial Intelligence (AI) and Machine Learning (ML) to address critical challenges posed by background noise and distortions. The study not only underscores the limitations of existing systems, especially in diverse real-world scenarios but also proposes a comprehensive solution through the development of an advanced audio fingerprinting algorithm. Building upon the foundations of the Dejavu Project, the research emphasizes the importance of robust data collection and analysis, simulating complex auditory environments with a diverse dataset. Signal processing techniques, including Fast Fourier Transform (FFT), spectrograms, and peak extraction, form the core of the Dejavu model, providing a nuanced understanding of audio signals amidst noise and distortions. The "constellation" concept, fingerprint hashing, and the intricate process of song learning and recognition contribute to the algorithm's sophistication. The research showcases impressive performance metrics, with recognition accuracy reaching 100% with just 5 seconds of audio input. The system's predictable matching speed, as demonstrated in the analysis of "Get Lucky" by Daft Punk, emphasizes its efficiency. However, the study also acknowledges the trade-off between storage and speed, with fingerprints occupying a substantial amount of space. This highlights a critical consideration for practical implementation. In essence, the research lays a solid foundation for advancing the adaptability of audio fingerprinting technology, opening avenues for more secure, efficient, and user-friendly applications. By integrating AI and ML, this breakthrough sets a new standard for accuracy in audio content identification, promising transformative implications for digital audio management in various sectors, including media, entertainment, and security